\documentclass[twocolumn,twocolappendix]{aastex631}
\usepackage{comment}
\usepackage{chemformula}
\usepackage{booktabs}
\usepackage{rotating}
\usepackage{mathtools}
\usepackage{graphicx}
\usepackage{soul}
\usepackage{hyperref}

\shorttitle{Binding energies and spectra of S-chains}
\shortauthors{J. Perrero et al.}
\graphicspath{{./}{figures/}}

\begin{document}

\title{Binding energies and vibrational spectral features of S$_n$ species on amorphous water-ice mantles: a quantum mechanical study}

\correspondingauthor{Albert Rimola, Piero Ugliengo}

\author[0000-0003-2161-9120]{Jessica Perrero}
\affiliation{Departament de Qu\'{i}mica, Universitat Aut\`{o}noma de Barcelona, Bellaterra, 08193, Catalonia, Spain}
\affiliation{Dipartimento di Chimica and Nanostructured Interfaces and Surfaces (NIS) Centre, Universit\`{a} degli Studi di Torino, via P. Giuria 7, 10125, Torino, Italy}

\author[0000-0003-0833-4075]{Leire Beitia-Antero}
\affiliation{AEGORA Research Group - Joint Center for Ultraviolet Astronomy, Universidad Complutense de Madrid, Plaza de Ciencias 3, 28040 Madrid, Spain}
\affiliation{Departamento de Estad\'{\i}stica e Investigaci\'on Operativa, Fac. de CC. Matem\'aticas, Plaza de Ciencias 3, 28040 Madrid, Spain}

\author[0000-0001-6317-6343]{Asunción Fuente}
\affiliation{Centro de Astrobiología (CSIC/INTA), Ctra.de Torrejón a Ajalvir km 4, E-28806 Torrejón de Ardoz, Spain}

\author[0000-0001-8886-9832]{Piero Ugliengo}
\affiliation{Dipartimento di Chimica and Nanostructured Interfaces and Surfaces (NIS) Centre, Universit\`{a} degli Studi di Torino, via P. Giuria 7, 10125, Torino, Italy}
\email{piero.ugliengo@unito.it}

\author[0000-0002-9637-4554]{Albert Rimola}
\affiliation{Departament de Qu\'{i}mica, Universitat Aut\`{o}noma de Barcelona, Bellaterra, 08193, Catalonia, Spain}
\email{albert.rimola@uab.cat}

\begin{abstract}
In the denser and colder regions of the interstellar medium (ISM), gas-phase sulfur is depleted by two or three orders of magnitude with respect to its cosmic abundance. Thus, which species are the main carriers of sulfur is an open question. Recent studies have proposed S$_n$ species as potential sulfur reservoirs. Among the various sulfur allotropes, the most stable one is the S$_8$ ring, detected in asteroid Ryugu and Orgueil meteorite. Shorter species, namely S$_3$ and S$_4$, have been found in comet 67P/C-G, but their presence in the ISM remains elusive. In this study, we compute the binding energies (BEs) of S$_n$ (n=1--8) species on an amorphous water ice surface model and analyze their infrared (IR) and Raman spectral features to provide data for their identification in the ISM. Our computations reveal that these species exhibit lower BEs than previously assumed, and their spectral features experience minimal shifts when adsorbed on water ice because of the weak and non-specific S$_n$/ice interactions. Furthermore, these species display very low IR band intensities and, therefore, very accurate instruments operating in the mid-infrared range are required for detecting the presence of these species in dense interstellar environments.

\end{abstract}

\keywords{Unified Astronomy Thesaurus concepts: Surface ices (2117) --- Interstellar dust (836) --- Interstellar molecules (849) --- Dense interstellar clouds (371) --- Interstellar medium (847) --- Solid matter physics (2090) --- Interstellar dust processes (838) --- Computational methods (1965)}

\section{Introduction} \label{sec:intro}

Sulfur chemistry has garnered significant interest in astrochemistry, particularly since sulfur depletion in the interstellar medium (ISM) was first acknowledged in the 1970s  \citep{spitzer1975ultraviolet,gondhalekar1985}.
Although the abundance of sulfur in the gas phase of the diffuse medium is consistent with the cosmic value, [S]/[H] = 1.5$\times$10$^{-5}$ cm$^{-3}$ \citep{jenkins2009}, more evolved environments such as translucent clouds and dense clouds are characterized by an environment-dependent depletion of one or two orders of magnitude \citep{fuente2023}. Initially, the sulfur missing from the gas phase was expected to yield H$_2$S on the icy grain mantles \citep{caselli1994}, similar to the freezing of oxygen atoms yielding H$_2$O. However, the abundance of S-bearing species observed in the ice mantle (OCS, and tentatively SO$_2$) and the upper limits estimated for H$_2$S, account for less than 5\% of the sulfur budget in the solid phase \citep{boogert2022,mcclure2023}. For this reason, the main reservoirs of sulfur are still unknown, and its chemistry is being extensively investigated.

Sulfur is known to present different molecular allotropes, forming chains and rings up to 20 atoms, as it easily tends to react with itself even in a diluted medium. In the last 20 years, the interest in the S$_n$ species (with 2 $\le$ n $\le$ 8) has grown exponentially in the field of astrochemistry, after being first proposed as S-reservoirs \citep{wakelam2004}. Several experiments have shown the production of these species, together with H$_2$S$_n$ (with n $\ge$ 2), after photoprocessing of H$_2$S-containing ices \citep[e.g.][]{ferrante2008, jimenez2011, jimenez2014, cazaux2022}. When including proper reaction pathways, astrochemical models also predict their formation \citep{laas2019}. In particular, when considering cosmic-ray-driven radiation chemistry and fast non-diffusive reactions for bulk radicals, pure sulfur allotropes are one of the main products \citep{shingledecker2020}.
So far, the only molecule characterized by a \ch{S\bond{single}S} bond was detected in the Horsehead nebula, a moderately UV-irradiated environment, where gas-phase S$_2$H was found \citep{fuente2017}. On the other hand, S$_3$ and S$_4$ have been detected in comet 67P/C-G (beside sulfides and sulfur oxides) by \cite{calmonte2016}. 
In the same source, \cite{altwegg2022} also detected ammonium sulfide, which is the most abundant salt of the comet. S$_n$ species have also been found in other bodies of the Solar system, such as the Ryugu asteroid and the Orgueil meteorite, which contain low and high abundances of S$_8$, respectively \citep{aponte2023}.

The formation mechanism of S$_n$ species is suspected to depend on the environment. Translucent cloud conditions are more likely to favor the formation of long S$_n$-species \citep{cazaux2022}, as \cite{ruffle1999} suggested that the negative grain charge distribution could be responsible for enhancing S$^+$ accretion on the core of the grains. On the other hand, laboratory experiments predict the formation of short S$_n$ species due to UV photoprocessing of H$_2$S ice, along with the formation of H$_2$S$_n$ species \citep{jimenez2011, jimenez2014}. The abundance ratio between the species that are formed depends on the shielding effect of the H$_2$S ice \citep{cazaux2022}. S$_n$ (n $>$ 4) species would be more refractory as \textit{n} increases, so that they could not desorb from the core of the grains in hot cores/hot corinos, but could be released in shock regions. On the other hand, S$_n$ (2 $\le$ n $\le$ 4) species could potentially be observed in less harsh environments, and hence the need for providing spectroscopic data for their identification.

With the launch of JWST in December 2021 and the beginning of the EUCLID mission in July 2023, numerous spectra will be collected, not only for gas-phase species, but also for interstellar ices, and the interpretation of the spectra will need both experimental and computational data.

From a theoretical point of view, there are several studies on the S$_n$ free molecules, which elucidate their geometrical features and electronic states \citep[e.g.,][]{raghavachari1990,mccarthy2004_s3,mccarthy2004_s4}. Their IR and Raman spectra have been recorded experimentally in the gas phase or with matrix isolation in solid argon technique, depending on which was more feasible for each species, which are collected in reviews \citep{eckert2003, trofimov2009}. Gas-phase S$_3$ and S$_4$ have also been characterized by means of rotational spectroscopy in the mm regions, providing valuable data for their detection and identification in the ISM \citep{mccarthy2004_s3,mccarthy2004_s4}.
However, to the best of our knowledge, no previous studies have been conducted for S$_n$ interacting with water ice. Thus, in this work, we focused on the adsorption of S$_n$ (n=1--8) species on interstellar water ices (here modeled as an extended periodic amorphous surface) with the aim to provide their binding energies (BEs) and vibrational spectroscopic (IR and Raman) properties computed at a quantum mechanical level. BEs are important parameters used in astrochemical models to describe the chemical evolution of planet-forming environments. Since BEs appear in exponential terms, their accuracy is mandatory to provide sensible results. Additionally, the perturbation of the vibrational IR and Raman bands of the S$_n$ species due to the adsorption at ice will be of great value for their experimental detection in the ISM ices (at least for the IR spectra). In Section \ref{sec:computational} we describe the methodology applied in the simulations, and in Section \ref{sec:results} we present the results and a discussion of their implications in the field of astrochemistry. Finally, in Section \ref{sec:conclusion} we summarize our conclusions.

\section{Methodology} \label{sec:computational}

\begin{figure}[htb]
    \centering
    \includegraphics[width=\linewidth]{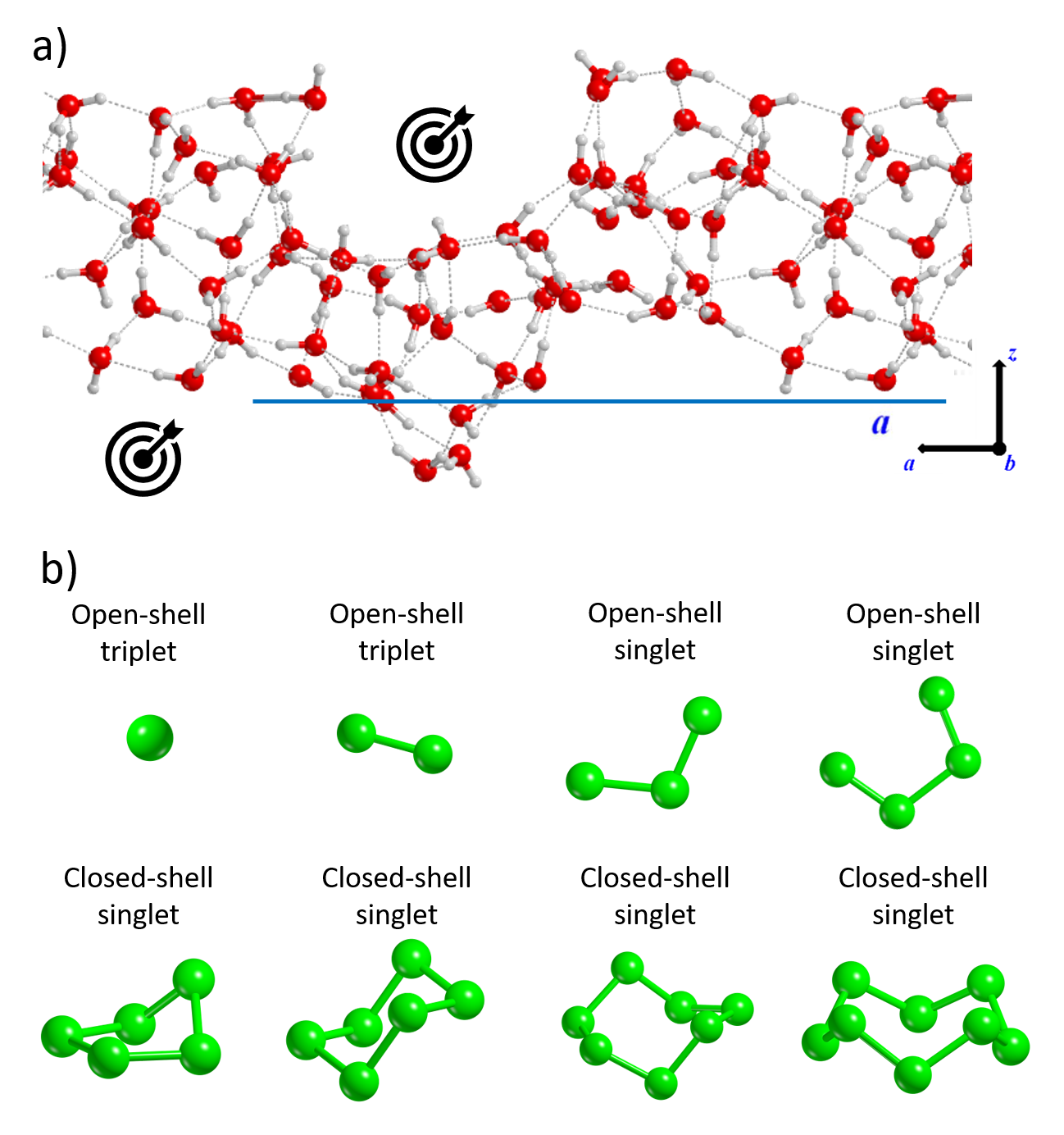}
    \caption{a) Amorphous periodic water ice model adopted in this work. The target icon indicates the binding sites considered for the adsorption of species from S to S$_8$. Colour code: red, oxygen; grey, hydrogen. b) Species characterized in this work with the corresponding electronic ground state. Species from S to S$_4$ are linear chains (top row), while species from S$_5$ to S$_8$ are cyclic rings (bottom row).}
    \label{fig:models}
\end{figure}

All the calculations were performed with the periodic \textit{ab initio} code \textsc{crystal17} \citep{dovesi2018}, which allows simulating systems from zero to three periodic dimensions (i.e., molecules, polymers, surfaces and bulks) and describe the atoms with Gaussian-type orbitals instead of plane waves (as commonly employed in periodic codes). Due to such a feature, our surfaces are slab models to which periodic boundary conditions (PBC) along two directions have been applied, presenting a finite thickness in the non-periodic direction.

After finding the most stable conformer and electronic state of each gas-phase S$_n$ species, we computed their adsorption on a periodic amorphous-like water ice surface (see Figure \ref{fig:models}). This model was designed by \cite{ferrero2020}, by joining three amorphous water clusters and applying the PBC, thus cutting the obtained bulk ice model along the (010) plane to yield a periodic surface. The resulting slab was characterized by edges and cavities, with binding sites possessing different adsorption strengths. The distinct morphologies of the upper and lower surfaces are responsible for the presence of a small electric dipole moment across the non-periodic direction. The unit cell has 180 atoms and its cell parameters are \textit{a}=20.275 \AA, \textit{b}=10.052 \AA, and $\gamma$=103.442° within the adopted density functional theory (DFT) level (see below). 

The geometry optimizations of the gas-phase S$_n$ species and their adsorption complexes were performed with the hybrid DFT BHLYP functional \citep{bhlyp-becke1993,lee1988}, supplemented with the D3(BJ) dispersion correction by \cite{grimme2010} and combined with the Ahlrichs-VTZ* (A-VTZ*) basis set, previously adopted in \cite{ferrero2020} and \cite{perrero2022}. The choice of this functional was justified by the need to treat both closed-shell and open-shell species at the highest cost-effective ratio. Thus, we relied on the known good performance of the B3LYP functional in similar previous publications \citep{ferrero2020,perrero2022} and chose to use a higher percentage of exact exchange (50\%) to properly describe open-shell species, resulting in the adoption of the BHLYP-D3(BJ)/A-VTZ* methodology.

The adsorption complexes were computed by relaxing both the internal atomic positions and the unit cell parameters. We selected two binding sites on the amorphous ice model in order to represent the two main binding sites that this model offers: i) the cavity region, and ii) a flat portion of the surface. In the previous works of \cite{ferrero2020} and \cite{perrero2022}, more binding sites were probed, with the aim to define BE ranges for a larger set of species which were, however, relatively small in size. In this case, since large non-polar species are involved (interacting mainly through weak non-specific forces), we suppose these two binding sites to be representative of the range of interactions between the molecule and the ice surface. We expect larger BEs for species adsorbed in the cavity region rather than in the flat region because of the larger contact surface between the adsorbate and the ice water molecules (and hence larger interactions), in the former case.

The final binding energies, BE(0), were computed through the equation:
\begin{equation}\label{eqn:BE(0)}
BE(0) = BE - \Delta ZPE= - (\Delta E - BSSE) - \Delta ZPE
\end{equation}
where the interaction energy $\Delta$E = E$_{complex}$ - E$_{ice}$ - E$_{species}$ was corrected for the basis set superposition error (BSSE, arising from the finiteness of the basis set) and for the zero-point energy ($\Delta$ZPE = ZPE$_{complex}$ - ZPE$_{ice}$ - ZPE$_{species}$) computed at 0 Kelvin. The ice surface taken as reference is the one optimized after the desorption of the adsorbate, this way neglecting spurious effects in the BE(0)s due to an over-deformation of the ice structure induced by the presence of the adsorbate and caused by the finite thickness of the ice model. Indeed, our experience indicates that an amorphous ice surface, contrary to a rigid crystalline one, can dramatically reconstruct locally to better accommodate the adsorbate \citep{perrero2022, perrero2023}. The significant deformation observed is the result of fully relaxing the geometry of the systems (both internal atomic positions and unit cell parameters), which is compulsory to ensure that the adsorption complexes are actual minima stationary points, and hence to accurately compute IR and Raman spectra. Consequently, during the optimization process, the ice unit cell may contract or expand, contrary to what would happen on a real ice surface. To address this discrepancy, we opted to derive binding energies from simulations of desorption processes. Basically, we simulate a desorption process rather than an adsorption one, in which, on the latter the reference surface is the pristine one, while for the former it is the one left after desorption. The different terms that contribute to the BE(0) were analyzed and explained in the Appendix (see Section \ref{appendix:be}), and for their detailed explanation, we refer the reader to the works of \cite{ferrero2020} and \cite{perrero2022}.

Frequency calculations on the BHYLP-D3(BJ)/A-VTZ*-optimized geometries were run on the adsorption complexes with the twofold aim: i) confirming their nature of minima on the potential energy surface (PES), and ii) computing the IR and Raman spectra. The Hessian matrix was calculated considering only the displacements of the S$_n$ adsorbed species, while the rest of the system is kept at fixed geometry. We previously checked that the exclusion of water molecules did not have a significant influence on the fragment Hessian matrix and, therefore, on the final vibrational spectra. For each coordinate, two displacements along each of the three cartesian coordinates were computed, with a step of 0.001 \AA~ and a tolerance on the SCF energy of 10$^{-11}$ atomic units. Both the IR and Raman intensities were computed through a Coupled-Perturbed Hartree-Fock/Kohn-Sham approach \citep{pascale2004, zicovich2004}, which allows for a completely analytical calculation of Born charges, IR intensities, and dielectric and Raman tensors. Please, note that although we used a fragment to calculate the frequencies involving only the S atoms of the S$_n$ species, the IR and Raman intensities are unaffected by this approach. Indeed,  
for the IR spectra, the transition dipole moment governing their intensities is calculated through a wave function encompassing the whole system, which accordingly accounts for the polarization effects of the ice. Similarly for the Raman intensities, as the transition polarizability is calculated with the complete wave function.
The vibrational modes were classified through the analyzer implemented in \textsc{crystal17}. The software decomposes the motion of each couple of atoms into three components: the first along the two atoms; the second on the plane containing a third atom; the third out of the above-mentioned plane. Finally, it uses this information to classify the modes as stretching (S), bending (B) or other (O).
The IR and Raman spectra were calculated by raw superposition of Lorentzian peaks, with a FWHM of 8 cm$^{-1}$, applying a Lorentzian broadening to the peaks after computing their relative intensities \citep{maschio2013a,maschio2013b}.

\section{Results \& Discussion} \label{sec:results}

\subsection{Gas-phase S$_n$ species}

Numerous studies in the literature have delved into the various molecular allotropes of sulfur. These investigations focused not only on the structural features and electronic states of the S$_n$ species but also on their spectroscopic properties.
The electronic ground state of atomic sulfur and S$_2$ is a triplet (like atomic oxygen and O$_2$), while species with n $\geqslant$ 6 are closed-shell monocyclic rings, which have been characterized by X-ray crystallography and infrared spectroscopy \citep{eckert2003}.
However, the S$_3$, S$_4$ and S$_5$ systems have sparked debates regarding their configuration as either chains or rings, prompting extensive research \citep[e.g.,][]{meyer1972infrared,mccarthy2004_s3,mccarthy2004_s4,thorwirth2005}. Figure \ref{fig:models} shows the most stable structures and their electronic states found with our computational methodology. In the following, we provide a comparison between our results and literature data. Singlet and triplet state energies of each species can be found in Appendix \ref{appendix:species}.

In their quantum chemical investigation, \cite{raghavachari1990} showed that the most stable conformation for S$_3$ was a chain, in a singlet electronic state. Our calculations agree with this hypothesis, when an open-shell singlet electronic configuration is assumed. Moreover, we found that the second most stable configuration is the closed shell singlet spin state, closely followed by a ring structure with a closed-shell singlet electronic state.

For S$_4$, DFT calculations have revealed six possible isomers, involving both chains and rings, which can be detected under non-equilibrium conditions \citep{eckert2003}. 

The open chain structure with C$_{2v}$ symmetry and an open-shell singlet electronic state was reported as the most stable isomer \citep{mccarthy2004_s4}, which is in agreement with our calculations at BHLYP-D3(BJ) level of theory. 

S$_5$ is a minority species in sulfur vapor at all temperatures and pressures \citep{steudel2003}, and is the least characterized structure, for which no spectral information is available. Only one minimum can be found on the PES of S$_5$, corresponding to the envelope conformation, while the half-chair conformation is a saddle point \citep{cioslowski2001}. In our calculations, we find that the energy difference between the two conformations is 0.7 kJ mol$^{-1}$ at DFT level of theory, and the presence of an imaginary frequency for the half-chair conformation confirms that the envelope structure is the correct minimum. 

Rings as S$_6$ and larger can also be found in the solid phase. S$_6$ is a highly symmetric homocycle, the most stable conformer being the chair conformation, in analogy to cyclohexane.  S$_7$ is better known for existing in four bulk allotropic forms, while a ring with C$_s$ symmetry is the favored conformer in the gas phase. S$_8$ constitutes the most thermodinamically stable and thus the common form of sulfur, and has been determined to adopt a crown-shaped geometry in the gas phase \citep{raghavachari1990}.

In summary, the ground states of the species targeted in this study are open-shell triplet for S and S$_2$, open-shell singlet for the chains S$_3$ and S$_4$, and closed-shell singlet for the rings from S$_5$ to S$_8$, all in agreement with previous literature data. 

\subsection{ S$_n$/ice adsorption complexes and BEs}

\begin{figure}[htb]
    \centering
    \includegraphics[width=\linewidth]{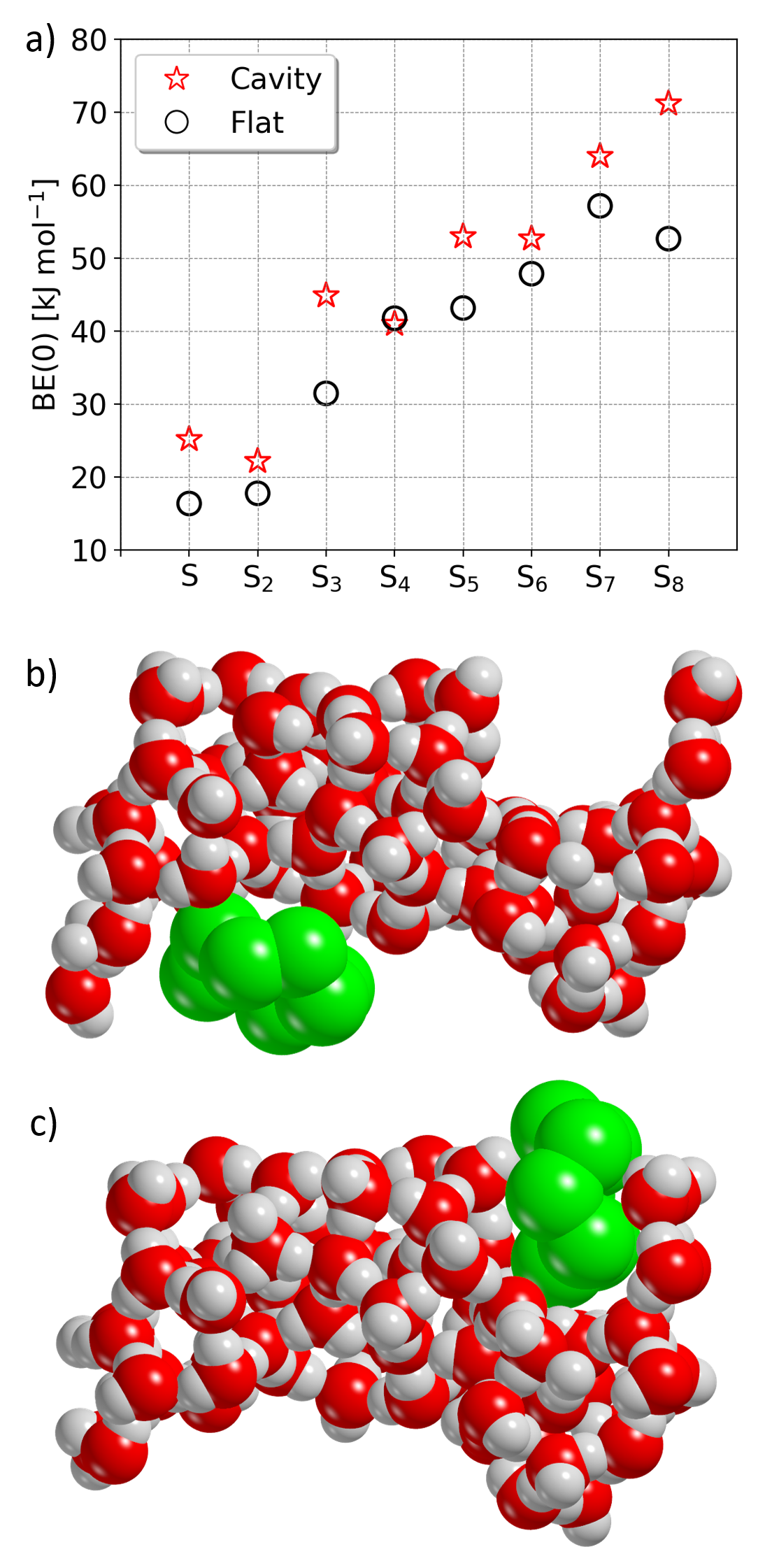}
    \caption{a) Binding energies (BE(0), in kJ mol$^{-1}$) of S-chains at BHLYP-D3(BJ)/A-VTZ* level of theory adsorbed onto the amorphous model (cavity site, star; flat site, circle). Adsorption geometries of S$_8$ onto the amorphous ice surface: b) flat region; c) cavity region.}
    \label{fig:bindings}
\end{figure}

For each S$_n$ species, we considered two adsorption complexes on the amorphous ice model, initially locating the species on regions either geometrically flat or in the small surface cavity (highlighted by the target icon in Figure \ref{fig:models}). Table \ref{tab:BE(0)} reports the computed BE(0) values and Figure \ref{fig:bindings} plots a comparative trend of the BE(0)s when the S$_n$ species are adsorbed in either the cavity or on the flat regions. The terms contributing to the BE(0) values are reported in Table \ref{tab:BEcontribution}.

The interaction between the S$_n$ species and the ice is almost entirely due to: i) electronic effects (contributed by electrostatic interaction, Pauli repulsion and orbital interaction due to charge transfer and polarization effects), and ii) London dispersion interactions. The percentage of the dispersion interactions contribution to each BE value (see Table \ref{tab:BEcontribution}) shows their dominance, with the exception of atomic sulfur, for which 40\% to 50\% of the $\Delta$E is due to the electronic contribution. For S$_2$ and S$_4$ cases, dispersion interactions even compensate for the slight repulsion from the pure electronic interactions $\Delta$E. Therefore, the binding energy is dominated by dispersion interactions, which are governed by the mutual polarizability of the interacting species (ice grain and the S$_n$ species). Indeed, S$_n$ species exhibit a highly polarizability but negligible electric multipoles providing small contribution to the electrostatic component of the BE. For the same reasons, also the charge transfer contribution is very small. Accordingly, we obtain dispersion contributions ranging 70--100\% of the total binding energy.

The BE(0) values reported in Table \ref{tab:BE(0)} indicate that S and S$_2$ are the two S-bearing species with the lower BE(0)s, obviously due to the more limited electronic and dispersion contributions. Remarkably, in the cavity region, atomic S interacts more strongly than S$_2$, as occurs for H and O compared to their corresponding (more inert and stable) diatomic H$_2$ and O$_2$ species \citep{ferrero2020atoms,ferrero2020,minissale2022}. 
From S$_2$ to S$_8$, the BE(0)s (and in particular the $\Delta$E$^*$ term, free from the deformation energies, see Table \ref{tab:BEcontribution}) increases with the number of S atoms, as each atomic S addition gives further interactions in the S$_n$/ice complexes.
That is, although dispersion forces are weak, the sum of several small contributions can result in large interaction energies, especially when the adsorbate has several atoms due to establishing a large contact area with the surface. This effect is responsible for the increase of the BE with the number of S atoms in the adsorbed species, even in the absence of local interactions such as the hydrogen bonds.\par

The morphology of the surface binding site affects the resulting BE(0). Indeed, in most cases, the BE(0) in the cavity region is larger than that on the flat region (see panel a) of Figure \ref{fig:bindings}). There is a good fit between the two sets of BE(0)s revealing an increase of about 18\% of the BE(0) when passing from flat to the cavity regions. As outlined before, the dispersion forces, responsible of the S$_n$/ice interactions, strongly depend on the contact area between the adsorbates and the surface water molecules, becoming larger for larger surface contact due to the higher polarizability effects onto the S$_n$ species. In the amorphous cavity regions, almost every S atom of the adsorbed species interacts with the surface water molecules, while on the flat region, only some atoms of the adsorbate are directly facing the water molecules of the surface, the rest being oriented toward the gas phase. This results in a higher weight of dispersion interactions for species adsorbed in the cavity region (see panels b and c of Figure \ref{fig:bindings}). The only exception is S$_4$ (with similar BE(0)s for the two regions), because the interaction of S$_4$ with the cavity is associated with a large deformation of the cavity surface region (listed in Table \ref{tab:BEcontribution}). 

The adsorption of large rings in the cavity means, to some extent, a structural rearrangement of the surface to fit the molecule in the cavity. This is reflected by the increment of the surface deformation energy ($\delta$E$_S$) with the increase of the number of S atoms, in particular for the S$_7$ and S$_8$ cases. In contrast, the deformation of the adsorbed species ($\delta$E$_{\mu}$) is of few kJ mol$^{-1}$, since they do not suffer from structural modifications due to their weak interactions with the ice.

The lateral interaction energy ($\delta$E$_L$) between adsorbates in adjacent cells (whose absolute value is mainly below 2 kJ mol$^{-1}$, see Table \ref{tab:BEcontribution}), is indicative of the adequate dimensions of the unit cells for the simulation of the freeze-out of an isolated S$_n$ species.

\begin{table*}[htb]
\centering
\caption{BE(0) in kJ mol$^{-1}$ for S$_n$ species adsorbed on the two binding regions of the amorphous ice model, computed at BHLYP-D3(BJ)/A-VTZ* level of theory. Values from the literature are listed for comparison. }\label{tab:BE(0)}
\begin{tabular}{l c c c c c c c c c }
BE(0) & Surface & \ch{S} & \ch{S2} & \ch{S3}  &   \ch{S4}  &   \ch{S5}  &   \ch{S6}  &   \ch{S7}  &   \ch{S8}  \\
\toprule
Flat (this work) & Comput., periodic H$_2$O ice &  16.4 & 17.8 & 31.5 & 41.8 & 43.2 & 47.9 & 57.5 & 52.7 \\
Cavity (this work) & Comput., periodic H$_2$O ice & 25.5 & 22.2 & 44.9 & 41.0 & 53.0 & 52.7 & 64.0 & 71.2 \\
\cite{perrero2022} & Comput., periodic H$_2$O ice & 13.1-23.3 & 8.6-20.4 & & & & & & \\
\cite{cazaux2022} & Exp., H$_2$S ice &  & 28.2$^a$ & 49.9$^a$ & 70.6$^a$ & 89.8$^b$ & 109.7$^b$ & 129.7$^b$ & 149.7$^b$ \\
\cite{penteado2017} & Exp. \& Comput. & 8.2 $\pm$ 4.1  & 16.6 $\pm$ 4.1  \\
\cite{wakelam2017} & Comput., 1 H$_2$O & 21.6 &  & & & & & &  \\
\cite{das2018} & Comput., (H$_2$O)$_4$ cluster & 11.9 & 13.7 & & & & & & \\
KIDA \citep{kida} & -- & 21.6$^c$ & 18.3$^d$ & & & & & & \\
UMIST \citep{umist} & -- & 9.1$^d$ & 18.3$^d$ & & & & & & \\
\bottomrule
\end{tabular}

\noindent
$^a$Values derived from laboratory experiments. $^b$Extrapolated values (see text for details). \\
$^c$Value from \cite{wakelam2017}. $^d$Values from \cite{hasegawa1992}.
\end{table*}

\begin{figure}[htb]
    \centering
    \includegraphics[width=0.95\linewidth]{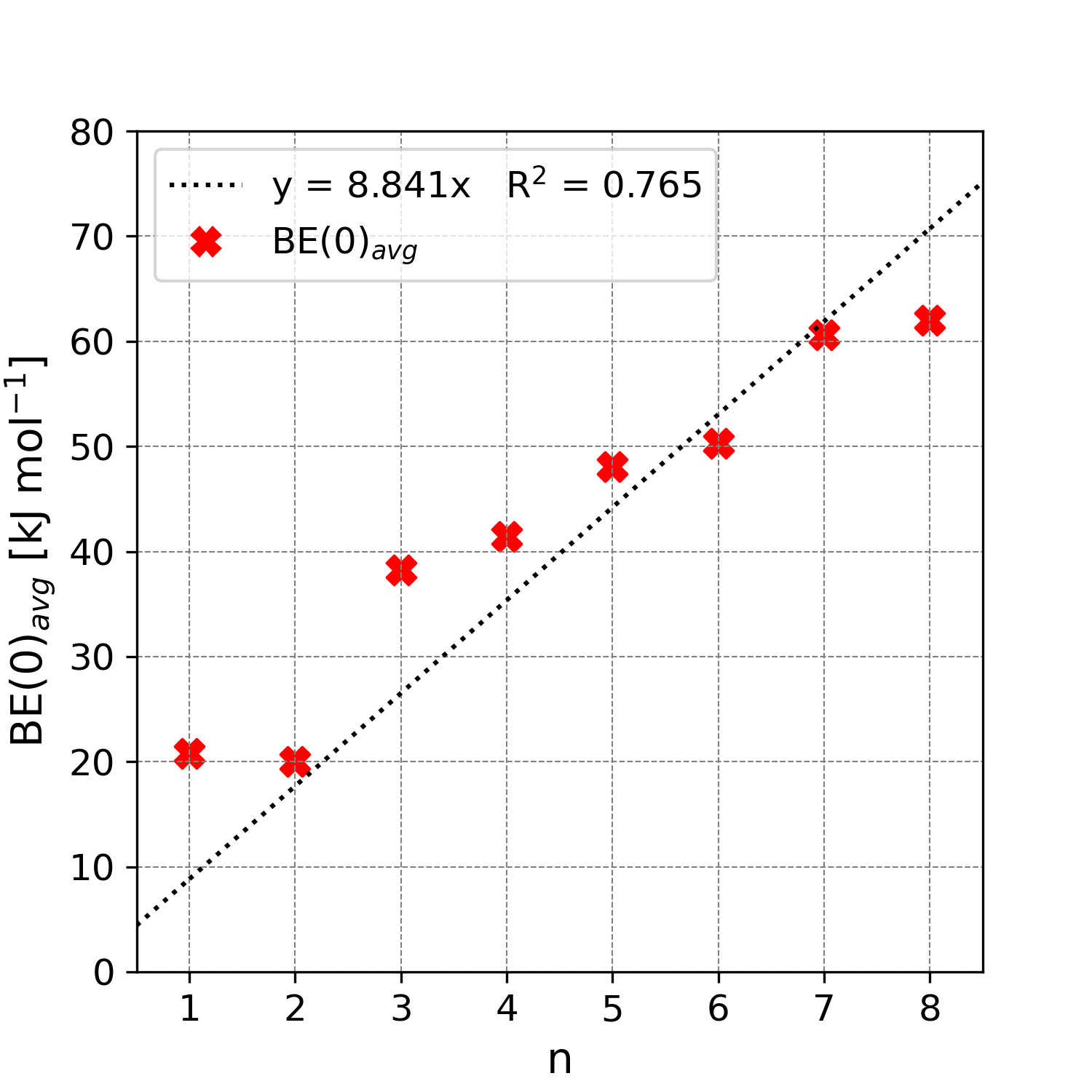}
    \caption{Correlation between the BE(0)$_{avg}$ and the number \textit{n} of atoms of each S$_n$ species. The linear fit does not seem appropriate to define the trend, as it appears that the increase of BE(0) happens in steps every couple of species. }
    \label{fig:correlation}
\end{figure}

In the literature, the only previous study on the BEs of species from S$_3$ to S$_8$ is based on an experimental work by \cite{cazaux2022} (see Table \ref{tab:BE(0)}). The BEs of S$_2$--S$_4$ are estimated from temperature programmed desorption (TPD) experiments, based on the empirical rule defined in \cite{luna2017}, while the BEs of S$_5$--S$_8$ are extrapolated based on the trend obtained for S$_2$--S$_4$.
However, the species were formed by H$_2$S photoprocessing, and therefore, their BE data are not relative to the adsorption on water ice, but rather the result of self interactions among the different sulfur species formed during the experiment. 
For this reason, the comparison between the BEs computed in this work and those estimated and/or extrapolated in \cite{cazaux2022} is not straightforward. A new set of experiments performed on water ice would be necessary to compare the data obtained from our calculations. From our own computed data, we plotted the average binding energy of each species, BE(0)$_{avg}$ = (BE(0)$_{cavity}$ + BE(0)$_{flat}$)/2, against the number of S atoms of the adsorbate (see Fig. \ref{fig:correlation}). Clearly, the increase in BE(0)s with the number of S atoms does not simply scale linearly. Rather, it seems that close couples (S$_1$--S$_2$, S$_3$--S$_4$, S$_5$--S$_6$ and S$_7$--S$_8$) exhibit almost the same BE(0). Although the first increase between pairs is steep (20 kJ mol$^{-1}$), the others are more moderate (10 kJ mol$^{-1}$). This seems to be due to the same number of S atoms facing the ice surface (three S atoms for S$_5$--S$_6$, four S atoms for S$_7$--S$_8$ cases), significantly changing the dispersive interactions. 
Even if quite rough for the above reasons, the linear best fit of Fig. \ref{fig:correlation} gives an increment of about 9 kJ mol$^{-1}$ to the BE(0)$_{avg}$ for each added S atom. This increment is approximately half of the value proposed in the literature, which was set to 20 kJ mol$^{-1}$ per S atom (see \cite{cazaux2022}). 
Therefore, we would like to raise awareness about the magnitude of the BEs reported in the literature, as its use in astrochemical modeling has a significant impact on the desorption rates.
To have deeper insights into this point (namely, the influence of an error in the BE on the desorption rate constant), we calculated the ratio between desorption rates, r$_{des}$ with an error $\delta$ on the BE, that is:

\begin{equation}\label{ratio}
r_{des} = \frac{k_{des}(BE)}{k_{des}(BE + \delta)} = \exp{\Bigl(\frac{\delta}{R T} \Bigr) }
\end{equation}

Note that r$_{des}$ does not depend on the BE itself. Larger errors will have a major impact at low temperatures, while they will be smeared out at high temperatures, bringing the $\delta$/RT term towards zero. Considering the limit of chemical accuracy, the error is $\delta$ = 4 kJ mol$^{-1}$, meaning that at T = 10 K such error implies a difference of 20 orders of magnitude on the k$_{des}$. At such low temperatures, however, thermal desorption of sulfur allotropes is negligible and the impact of our estimations in the gas chemistry would be moderate. In contrast, thermal desorption can be dominant in warmer regions, around young protostars. In these environments, dust temperatures are $>$ 50 K and some volatile compounds can be sublimated. At these temperatures, an error of $\delta$ = 4 kJ mol$^{-1}$ reduces the difference on the k$_{des}$ to $\sim$ 4 orders of magnitude, with a significant impact in the abundances of these species in gas phase. Thus, the impact of minor inaccuracies in the BE on the results of a chemical model depends on the specific conditions of the region under study. The same absolute error on different BE values will have the same impact on the desorption rate, in which such an impact depends in turn on the considered environment.

\begin{table*}[htb]
\centering
\caption{Contributions to the BE(0)s (in kJ mol$^{-1}$) of the adsorption complexes modeled on the flat region and in the cavity region of the amorphous ice model, computed at BHLYP-D3(BJ)/A-VTZ* level of theory. The equations showing how the terms add up are reported and explained in the Appendix \ref{appendix:be}. $\Delta$E is the bare interaction energy, splitted in the electronic $\Delta$E$_{elect}$ and dispersive contributions $\Delta$E$_{disp}$, the latter also in percentage; $\delta$E$_S$ is the deformation energy of the surface; $\delta$E$_\mu$ is the deformation energy of the adsorbate; $\delta$E$_L$ is the lateral interaction energy; $\Delta$E$^*$ is the deformation-free interaction energy; and the BSSE is the basis set superposition error.}
\label{tab:BEcontribution}
\begin{tabular}{l c c c c c c c c }
\hline
Flat Region &   \ch{S} & \ch{S2} & \ch{S3}  & \ch{S4}  & \ch{S5}  & \ch{S6}  & \ch{S7}  &  \ch{S8} \\ 
\toprule
$\Delta$E         & -18.3 & -23.3 & -39.3 & -53.0 & -50.9 & -60.2 & -68.0 & -63.7 \\
$\Delta$E$_{elect}$  & -7.1 & +0.5 & -9.4 & +0.9 & -20.1 & -1.4 & -12.1 & -8.7 \\
$\Delta$E$_{disp}$ (\%) & -11.2 (61) & -23.8 (102) & -29.9 (76) & -54.0 (102) & -30.9 (61) &  -58.8 (98) & -55.9 (82) & -55.0 (86) \\
$\delta$E$_S$     & 0.7 & 0.9 & 5.1 & 3.0 & 5.2 & 13.7 & 4.7 & 3.8\\
$\delta$E$_\mu$   & 0.0 & 0.3 & 3.3 & 9.0 & 0.5 & 0.5 & 0.4 & 1.2 \\
$\delta$E$_L$             & 0.0 & -0.3 & -0.1 & -0.1 & -0.4 & -0.6 & -1.0 & -1.5\\
$\Delta$E$^*$     & -19.0 & -24.2 & -47.6 & -65.0 & -56.1 & -73.8 & -72.1 & -67.2 \\
BSSE              & -1.9 & -4.3 & -6.0 & -8.9 & -5.8 & -10.5 & -8.5 & -9.5\\
\hline
\hline
Cavity Region & \ch{S} & \ch{S2} &  \ch{S3}  & \ch{S4} &  \ch{S5} & \ch{S6} & \ch{S7} &  \ch{S8}  \\
\toprule
$\Delta$E        & -27.6 & -28.3 & -53.8 & -51.5 & -63.4 & -64.6 & -78.3 & -87.7 \\
$\Delta$E$_{elect}$  & -14.0 & +1.7 & -15.2 & +0.8 & -11.3 & +0.9 & -6.3 & -2.8 \\
$\Delta$E$_{disp}$ (\%) & -13.6 (49) & -30.0 (106) & -38.7 (66) & -52.3 (102) & -52.1 (82) & -65.5 (101) & -72.0 (92) & -85.0 (97) \\
$\delta$E$_S$    & 5.0 & 1.6 & 2.3 & 10.7 & 8.3 & 6.4 & 10.3 & 9.7\\
$\delta$E$_\mu$  & 0.0 & 0.0 & 0.0 & 0.0 & 0.6 & 0.3 & 0.4 & 0.7 \\
$\delta$E$_L$            & 0.0 & 0.0 & -0.1 & -0.3 & -0.4 & -0.4 & -0.7 & -1.4\\
$\Delta$E$^*$    & -32.7 & -29.9 & -56.0 & -62.0 & -71.9 & -70.9 & -88.3 & -96.8 \\
BSSE             & -2.4 & -4.8 & -6.9 & -8.7 & -8.4 & -10.3 & -12.2 & -14.7\\
\bottomrule
\end{tabular}
\end{table*}

As for atomic S and S$_2$ molecule, \textit{ab initio} literature calculations have been performed to compute their BEs on interstellar ice mantles. For atomic S, \cite{wakelam2017} used one water molecule to simulate the ice mantle, while \cite{das2018} used a water tetramer as ice analogue for studying S and S$_2$ adsorptions. An estimate of the BE of S and S$_2$ was provided by \cite{penteado2017}, in a review collecting values from both theoretical and experimental works. Other values of the BEs are tabulated in UMIST and KIDA databases \citep{umist,kida}. We also contributed, in a recent work \citep{perrero2022}, to the BEs of S and S$_2$, while giving a range of BEs (computed at the M062X-D3/A-VTZ*//HF-3c level of theory) for each species on eight different binding sites available in the amorphous ice model of this work. Interestingly, the BEs computed in the present work for S and S$_2$, cover the upper and lower limits of the ranges provided in \cite{perrero2022}.
The computed BEs of S$_2$ were in agreement with previous literature values, while for atomic S the weakest tail of the calculated BE range did not include the even smaller values proposed by \citep{penteado2017} and by the UMIST database \citep{umist}.  
All the BE values reported in the literature are collected in Table \ref{tab:BE(0)}.

\subsection{Vibrational spectroscopic features}

\begin{figure}[htb]
    \centering
    \includegraphics[width=\linewidth]{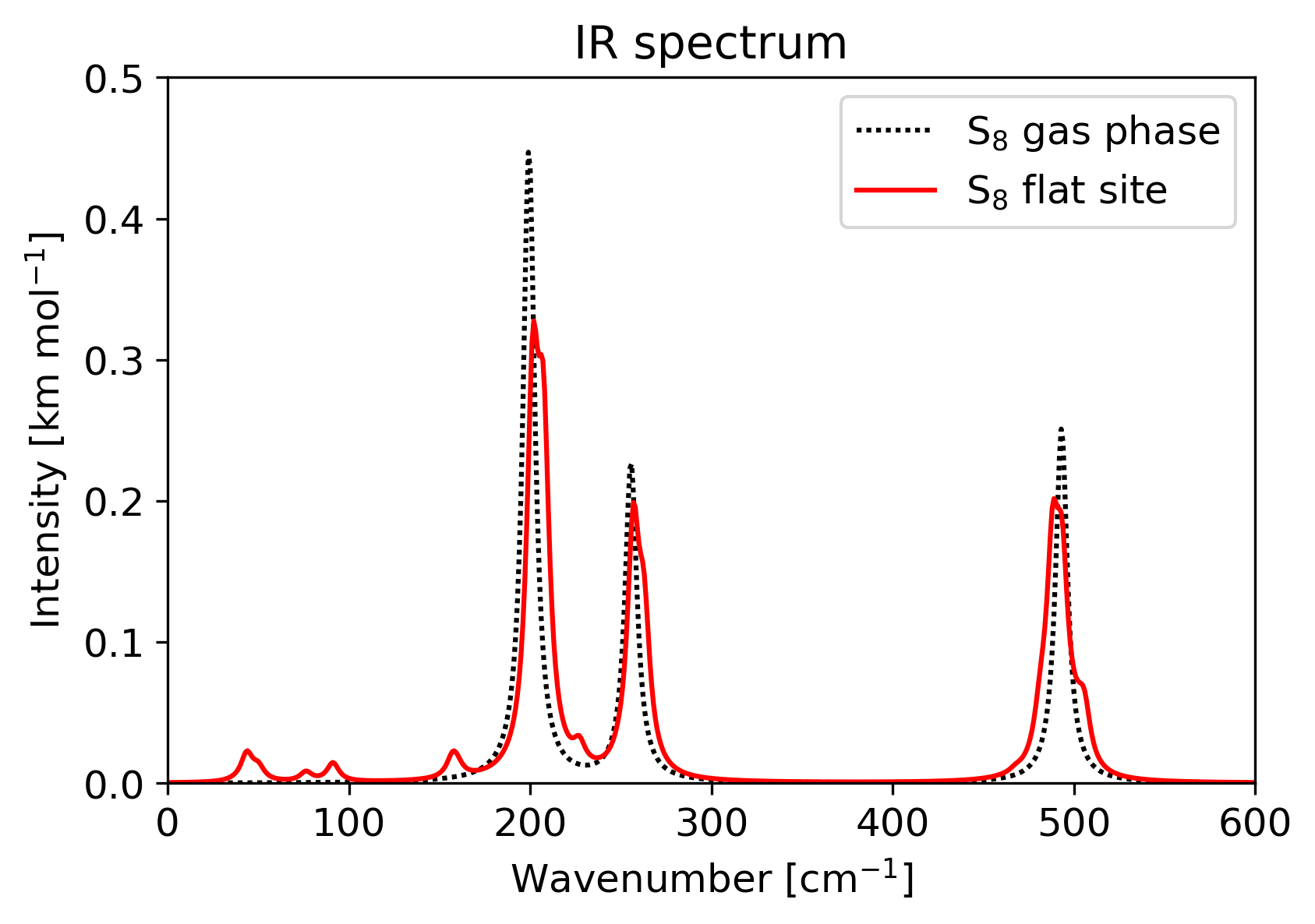} \\
    \includegraphics[width=\linewidth]{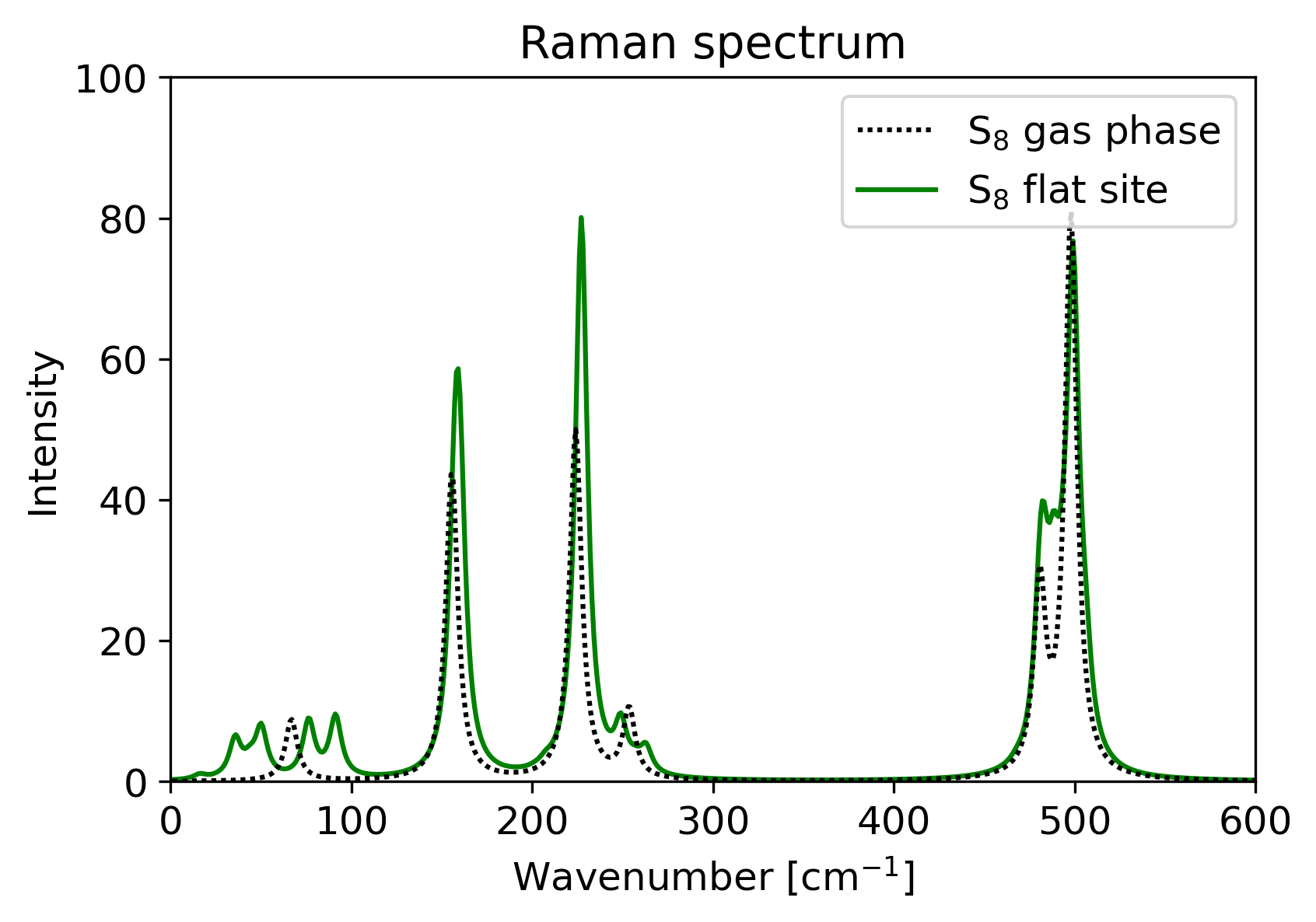}
    \caption{IR and Raman spectra of S$_8$ computed at BHLYP-D3(BJ)/A-VTZ* level of theory. Dotted lines: gas-phase species. Solid lines: adsorbed species. IR intensities are in km mol$^{-1}$, while Raman intensities are given in arbitrary units. 
    }
    \label{fig:spectra}
\end{figure}

Here, we focus on the calculation of the frequencies ($\nu$) of the S$_n$ species (for n=2--8) as gas-phase and as adsorbed complexes, for IR and Raman spectra.  Table \ref{tab:spectra} shows the whole set of computed vibrational frequencies for the considered cases, while Figure \ref{fig:spectra} shows the comparison between the largest S$_8$ molecules in gas-phase and when adsorbed on the flat region. 

Several studies have measured experimentally or simulated theoretically the spectra of these species, both in the gas phase or in the condensed phase. 
The reviews of \cite{eckert2003} and \cite{trofimov2009} are particularly useful for gathering detailed information. The main vibrational features available in the literature are reported in Table \ref{tab:spectra}.
It is worth noting that the comparison between the frequencies reported in the literature and those computed in this work is not always satisfying. For example, \cite{lenain1988} reports a stretching vibration for S$_2$ of 715 cm$^{-1}$, while our calculations give a value of 747 cm$^{-1}$. 
For S$_3$, \cite{brabson1991} and \cite{picquenard1993} give 281 cm$^{-1}$ (Raman) for the bending, and 581 cm$^{-1}$ (Raman) and 680 (IR) cm$^{-1}$ for the symmetric and antisymmetric stretching, respectively, while we find values of 254, 520 and 639 cm$^{-1}$. 
We can immediately see that the differences are large, even considering that highly accurate quantum-chemical calculations have been performed, in agreement with what is reported in the literature \citep{steudel2003}.

It is worth noting that no scaling was applied to our computed frequencies. Nevertheless, the absolute prediction of the vibrational frequencies is not the focus of this work; we are rather interested in highlighting the frequency shifts of the S$_n$ species vibrational modes with respect to the gas-phase values due to their interaction with the ice. 

As mentioned above, the interaction between the S$_n$ species and the water ice surfaces is almost totally governed by London dispersion forces, which are very weak forces based on permanent dipole -- induced dipole moments, and therefore are non-specific and non-directional (at variance with the H-bonding). In our case, the dispersion contribution arises from the interaction between the permanent dipole moment of the water surface molecules and the temporary or induced dipoles of the S$_n$ adsorbates. The fact that dispersion forces are non-specific and do not establish vibrational coupling with the water surface causes the adsorption of S$_n$ species to be either only slightly perturbed or to have a null affect to the vibrational modes of the adsorbates. Consequently, the infrared spectra are almost unperturbed with respect to the free S$_n$ species.
Additionally, the shifts observed when comparing gas-phase and adsorbed species vibrational features further decrease with the molecular size of the adsorbate.
This is in agreement with the decreasing contribution of the electronic component of the BE(0) with the molecular size and the increasing role of the non-specific London type interactions. In general, we notice that both IR and Raman peaks suffer a small hypsochromic shift, due to a slight increase in the bond strength. We notice such an effect on S$_3$, whose bond lengths change from 1.951 \AA   \ in the gas phase to 1.901 \AA  \ and 1.930 \AA \ for the adsorbed molecule on the flat and cavity region of the ice. Such variations correspond to a hypsochromic shift of $\Delta \nu$ = 70 cm$^{-1}$ on the asymmetric stretching vibration (the only active IR mode). However, such a strong variation is observed only for S$_3$ (and S$_4$, see below), while for the other species the caused change is much smaller (about $\Delta \nu$ = 0--10 cm$^{-1}$) . S$_2$ experiences no shift due to adsorption, as evidenced by the unaltered bond length (1.899 \AA) and is consistent with its weak interaction with ice. While the S$_2$ stretching peak is active in Raman spectroscopy, in the IR spectra it is possible to distinguish between the gas-phase and the adsorbed cases. For symmetry, no bands are observed in the gas-phase, while the interaction with the ice slightly activates this mode. The same phenomenon has been documented for CH$_4$ and NH$_4$SH, whose IR spectral features are affected by the presence of an amorphous water ice matrix interacting with the adsorbate \citep{escribano2014,hudson2015}. Finally, for S$_4$, an interesting aspect appears: while there is no significant difference between the gas-phase and on the cavity region spectra, this is juxtaposed by a large change in the vibrational features on the flat region. That is, $\Delta \nu$ of the first two bending modes is 20--30 cm$^{-1}$, and $\Delta \nu$ $>$ 100 cm$^{-1}$ for the third vibrational mode. The explanation can be found in the geometrical features of the molecule: for all the cases (gas-phase and on both ice regions) the optimized conformer of S$_4$ falls into the C$_{2v}$ symmetry.  The gas-phase S-S bond length of 2.201 \AA \ shorten to 2.193 \AA  \ when adsorbed on the cavity, with a corresponding hypsochromic shift of $\Delta \nu$ = 20-30 cm$^{-1}$. Even more so, when adsorbed on the flat region, the S--S bond length decreases to 2.056 \AA, with a large hypsochromic shift of its stretching frequency ($\Delta \nu$ $>$ 100 cm$^{-1}$). 

In principle the present outcome may guide the interpretation of the JWST spectra detection of interstellar ice, at least for what concerns the IR spectra. However, the absolute intensity of each IR vibrational mode is an important parameter as it regulates the chance of being observed. Our results show many IR peaks with almost negligible intensities (which become smaller as the length of the S-chain increases), as the dipole changes during vibration are almost negligible for the homo-nuclear S$_n$ rings. It is a pity that Raman spectra of the interstellar matter cannot, unfortunately, be observed with the present technology as the Raman intensities, depending on the polarizability of the molecule, are rather high. Indeed, sulfur is a large and easily polarizable atom, imparting high polarizability to the S$_n$ species. For the time being, Raman can be adopted in terrestrial experiments of the kind reported by \cite{cazaux2022} to characterize the S-bearing species.

Ices in interstellar environments are primarily identified by their vibrational transitions in the near-to-far IR. In general, the 1--3 $\mu$m wavelength region contains combination and overtone modes, the 3--6 $\mu$m  region comprehends the stretching vibrations, the 6--30 $\mu$m region the bending and libration vibrations, while the 25--300 $\mu$m region represents torsional and intermolecular (lattice) modes \citep{boogert2015}. 

The frequencies covered by the vibrational modes of the investigated S$_n$ species span the 150--750 cm$^{-1}$ range, corresponding to a wavelength range of 13.3--66.6 $\mu$m. Above 30 $\mu$m the availability and capability of instrumentation are limited, e.g. the JWST is designed to observe in the wavelength range from 0.6 to 27 $\mu$m \citep{rauscher2005jswt}. 
In addition, both amorphous and crystalline olivine and forsterite exhibit bending and torsional modes within the 14--24 $\mu$m region \citep{zamirri2020, boogert2015}, and libration modes of water cover the region around 10--12 $\mu$m \citep{gibb2000,mcclure2023}. 

Therefore, the forest of vibrational bands located in that region of the spectra, along with the weak intensities of the vibrational S$_n$ peaks, may render their IR detection rather difficult.
When moving to terrestrial laboratory experiments, the seminal work of \cite{raman_sulfur} suggested that Raman spectroscopy is ideal for the characterization of sulfur allotropes, due to the ability to distinguish and assign several active vibration modes, as also shown in Table \ref{tab:spectra} from the computed ones. Thus, Raman bands simulated here can serve as proxies for the detection of the S$_n$ species adsorbed on the ices in terrestrial laboratory experiments. This would be the case, for instance, of the stretching modes around 500 cm$^{-1}$, which are all active in Raman while invisible in IR.

\begin{table*}[htb]
    \centering
    \caption{Vibrational frequencies in cm$^{-1}$ computed at full BHLYP-D3(BJ)/A-VTZ* level. The first seven row are dedicated to computed values, while the last three column are dedicated to literature values. The second row reports the frequencies computed for the gas-phase molecule, while the third and the fourth columns report the ones of the adsorbate in the cavity and on the flat region of the amorphous model. The peaks are classified as stretching (S), bending (B), and other (O, representing different types of torsion), and it is indicated if they are IR or Raman active. The literature values are taken from \cite{eckert2003} with the exception of S$_4$, for which we refer to the work of \cite{picquenard1993} $^{(a)}$. For the sake of clarity, only strong bands for each species are reported. Origin of the spectra: S$_2$ was recorded in the gas phase; S$_3$ from sulfur vapors and matrix isolation in solid argon; S$_4$ in saturated and unsaturated sulfur vapor; solid S$_6$, $\alpha$-S$_7$, and $\alpha$-S$_8$ were used to record their spectra. For a complete list of all the features, we refer the reader to \cite{eckert2003} and \cite{trofimov2009}.}
    \label{tab:spectra}
    \begin{tabular}{l c c c c c c | c c c }
         Specie & Gas-phase & Cavity Region & Flat Region & Type & IR & Raman & Tabulated Freq & Type & Source\\
     \toprule
         S$_2$ & 747  & 745  & 746  & S & & X &  715 & S & Raman\\
         
         S$_3$ & 254 & 267 & 264 & B & & X &  281 & B & Raman\\
                & 520 & 518 & 556& S & & X & 581 & S sym & Raman\\
                & 639 & 643 &  711 & S & X & X & 680& S antsym & IR\\

         S$_4$ & 140 & 151 & 171 & B & & X & & &\\  
            & 310 & 315 & 330 &  B & & X & 303$^a$ & B & Raman\\ 
            & 322 & 336 & 458 & B & X & X & & &\\
            & 656 & 655 & 669 & S & X & & 575$^a$  & S & Raman\\
            & 662 & 665 & 670/681 & S & & X & 680$^a$/683 & S &Raman/IR\\
        
         S$_5$ & 245 & 254 & 253 & O & X & X & & &\\ 
             & 306 & 308 &307 & B & X & X & & & \\
             & 416 &426/436 &427 & S & &X & & & \\
              & 423 &  426 &  423 &  S & X &  & & & \\        
             & 460 & 467 & 466& S & X & X & & & \\
              & 525 &  516 &  524 &  S &  X &  & & & \\      
             & 528 & 516/535 & 524 & S & &X & & & \\
               
          S$_6$ & 167 & 175 & 176&  O & X & & 180 & O & IR\\
                & 215 & 219 & 219&  O& X(no gas) & X & & & \\
                & 274  & 278 & 277& O& & X & & & \\
                 &329 & 333 & 333 & O & X& & 313 &B & IR\\
                 &-  &441 & 444 & S & X&  & & & \\
                & 486 & 490 & 490& S& X & X & 463 &S & IR \\
                & 504 & 505 & 504& S& & X  & & & \\
            
         S$_7$ & 160 & 169 & 165 & O & &X & & & \\  
            & 181 & 186 & 187 & O & X(no gas) & X & & & \\
            & 243 & 251 & 247 & B & X & X & 235 & B & IR\\
            & 285 & 290 & 288 & B & X & & 270 & B  & IR \\
            & 300 & 298 & 303 & B & & X  & & & \\
            & 411 & 410 & 409 & S & X & X & 400 & S  & IR\\
            & 451 & 464 & 459 & S & X & X & 480 & S & IR\\
            & 506 & 504 & 507 & S & X & X & 513 & S & IR\\
            & 536 & 527 & 526 & S & & X & & & \\
           
          S$_8$  & & & & & & & 90 & O & IR/Raman \\
            & 155 & 162 & 159 & B & &X & 150-250 & B & IR/Raman \\
            & 199 &  207 &  202 &  O & X &  & & &\\
            & 224 & 230 & 227 & O & & X & & &\\
            & 253 & 255 & 249 & O & & X & & &\\
            & 256 & 261 & 257 & O & X & & & &\\
            & 481 & 488 & 482 & S & & X & 410-480 & S & IR/Raman\\
            & 493 & 494 & 489 & S & X & & & &\\
            & 498 & 501 & 499 & S & & X & & &\\         
            \bottomrule
    \end{tabular}
    
\end{table*}

\section{Conclusions} \label{sec:conclusion}
In this work, we computed the zero point corrected binding energies (BE(0)s) and the vibrational features (both IR and Raman spectra inclusive of absolute intensities) of the adsorption complexes of S$_n$ species (n=1--8) on an amorphous water ice surface model by means of quantum chemical simulations. The ice model was already proposed by some of us \cite{ferrero2020,perrero2022} to study the interaction of many different molecules of astrochemical interest. 

We adopted \textit{ab initio} periodic calculations based on the BHLYP-D3(BJ)/A-VTZ* model chemistry for both the geometry optimization and the computation of IR and Raman spectra in the harmonic approximation (for frequencies and intensities). To simplify the modeling due to the large size of the S$_n$ species, we focused on two different regions of the amorphous ice model. The first exhibits a cavity, mimicking some ice porosity in which the contact between the adsorbate and the ice surface is at its maximum. The second is a flat region, which represents a portion of dense icy dust grains. This limits the number of BEs (mainly two) that we obtain, but we expect these values to well represent the upper and lower limits of the range that one would obtain by conducting a broader sampling of binding sites. 
The BE(0)s of the S$_n$ species resulted about 18\% higher for the cavity region than for the flat one, as expected due to the non-specific nature of the interactions. These two situations likely represent the tail limits of a complete BE(0)s distribution that can be computed on much larger (and not feasible with our computational facilities) icy grain models. 
Currently, a few chemical models exist where more than one BE value for each species can be included. Moreover, several works are focusing on the computation of BEs distribution histograms, e.g., \cite{bovolenta2022, tinacci2022, tinacci2023}, as more representative of the variety of binding sites that a real amorphous ice mantle could offer. A distribution of values with a statistical meaning allows determining a mean BE and a standard deviation for each species. However, if we were to suggest a single BE value to be adopted in chemical models, the average between the two BEs computed for each species should be a reasonable choice. Nonetheless, we are aware that a more meticulous sampling of the ice model would be necessary in order to provide statistically meaningful data.

We found that the BE(0) values of the S$_n$ species increase with the number of S atoms, and that their interaction with the ice is mainly driven by dispersion-like (London) interactions. Only for the single S atom, a major electronic effect contribution was detected.
We found only a rough linear correlation between the averaged BE(0) values of the two surface regions and the number of sulfur atoms, with an increment by each S atom of about 9 kJ/mol, which may be useful to guess the BE(0) for even larger sulfur polymorphs by simple addition. A deeper analysis revealed that close couples (S$_1$--S$_2$, S$_3$--S$_4$, S$_5$--S$_6$ and S$_7$--S$_8$) exhibit almost the same BE(0).
From the analysis of the IR and Raman spectra, it is not possible to easily distinguish between gas-phase and adsorbed species, given that the perturbations suffered by the vibrational features of the adsorbed species are small and, overall, their amount becomes smaller with the increasing size of the S$_n$ species. Thus, the species exhibiting significative vibrational frequency perturbations upon adsorption are S$_n$ with n=2--4. The intensities of the IR and Raman peaks suggest Raman spectroscopy (due to the large polarizability of sulfur and its allotropic forms) to be the technique of election for the detection of S$_n$ species in terrestrial laboratory experiments as Raman cannot, unfortunately, be used for studying the ISM spectral features.

\section{Acknowledgments} 
This project has received funding within the European Union’s Horizon 2020 research and innovation programme from the European Research Council (ERC) for the projects ``Quantum Chemistry on Interstellar Grains” (QUANTUMGRAIN), grant agreement No. 865657, and ``The trail of sulfur: from molecular clouds to life” (SUL4LIFE), grant agreement No. 101096293. 
The Italian Space Agency for co-funding the Life in Space Project (ASI N. 2019-3-U.O), the Italian MUR (PRIN 2020, Astrochemistry beyond the second period elements, Prot. 2020AFB3FX) are also acknowledged for financial support. Authors (J.P. and P.U.) acknowledge support from the Project CH4.0 under the MUR program ``Dipartimenti di Eccellenza 2023-2027” (CUP: D13C22003520001).
The Spanish MICINN is also acknowledged for funding the projects PID2021-126427NB-I00 (A.R.), PID2019-106235GB-I00 (A.F.), PID2020-116726RB-I00 (L.B.-A.), and CNS2023-144902 (A.R.).
The authors thankfully acknowledge RES resources provided by BSC in MareNostrum to activities QHS-2022-3-0007 and QHS-2023-2-0011, and the supercomputational facilities provided by CSUC.

Supplementary Material consisting of (i) the fractional coordinates of DFT adsorption structures optimized for the amorphous ice models, and (ii) images of the computed IR and Raman spectra, is available at \url{https://zenodo.org/records/10650677}.

\bibliography{my_biblio}{}

\begin{thebibliography}{}
\expandafter\ifx\csname natexlab\endcsname\relax\def\natexlab#1{#1}\fi
\providecommand{\url}[1]{\href{#1}{#1}}
\providecommand{\dodoi}[1]{doi:~\href{http://doi.org/#1}{\nolinkurl{#1}}}
\providecommand{\doeprint}[1]{\href{http://ascl.net/#1}{\nolinkurl{http://ascl.net/#1}}}
\providecommand{\doarXiv}[1]{\href{https://arxiv.org/abs/#1}{\nolinkurl{https://arxiv.org/abs/#1}}}

\bibitem[{Altwegg {et~al.}(2022)Altwegg, Combi, Fuselier, Hänni, De Keyser,
  Mahjoub, Müller, Pestoni, Rubin, \& Wampfler}]{altwegg2022}
Altwegg, K., Combi, M., Fuselier, S.~A., {et~al.} 2022, MNRAS, 516, 3900,
  \dodoi{10.1093/mnras/stac2440}

\bibitem[{{Anderson} \& {Loh}(1969)}]{raman_sulfur}
{Anderson}, A., \& {Loh}, Y.~T. 1969, Can. J. Chem., 47, 879,
  \dodoi{10.1139/v69-145}

\bibitem[{{Aponte} {et~al.}(2023){Aponte}, {Dworkin}, {Glavin}, {Elsila},
  {Parker}, {McLain}, {Naraoka}, {Okazaki}, {Takano}, {Tachibana}, {Dong},
  {Zeichner}, {Eiler}, {Yurimoto}, {Nakamura}, {Yabuta}, {Terui}, {Noguchi},
  {Sakamoto}, {Yada}, {Nishimura}, {Nakato}, {Miyazaki}, {Yogata}, {Abe},
  {Okada}, {Usui}, {Yoshikawa}, {Saiki}, {Tanaka}, {Nakazawa}, {Tsuda},
  {Watanabe}, {Hayabusa2-initial-analysis SOM Team}, \&
  {Hayabusa2-initial-analysis core Team}}]{aponte2023}
{Aponte}, J.~C., {Dworkin}, J.~P., {Glavin}, D.~P., {et~al.} 2023, EPS, 75, 28,
  \dodoi{10.1186/s40623-022-01758-4}

\bibitem[{Becke(1993)}]{bhlyp-becke1993}
Becke, A.~D. 1993, J. Chem. Phys., 98, 1372, \dodoi{10.1063/1.464304}

\bibitem[{Boogert {et~al.}(2022)Boogert, Brewer, Brittain, \&
  Emerson}]{boogert2022}
Boogert, A., Brewer, K., Brittain, A., \& Emerson, K. 2022, ApJ, 941, 32,
  \dodoi{10.3847/1538-4357/ac9b4a}

\bibitem[{Boogert {et~al.}(2015)Boogert, Gerakines, \& Whittet}]{boogert2015}
Boogert, A.~A., Gerakines, P.~A., \& Whittet, D.~C. 2015, ARA\&A, 53, 541,
  \dodoi{10.1146/annurev-astro-082214-122348}

\bibitem[{{Bovolenta} {et~al.}(2022){Bovolenta}, {Vogt-Geisse}, {Bovino}, \&
  {Grassi}}]{bovolenta2022}
{Bovolenta}, G.~M., {Vogt-Geisse}, S., {Bovino}, S., \& {Grassi}, T. 2022,
  \apjs, 262, 17, \dodoi{10.3847/1538-4365/ac7f31}

\bibitem[{Brabson {et~al.}(1991)Brabson, Mielke, \& Andrews}]{brabson1991}
Brabson, G.~D., Mielke, Z., \& Andrews, L. 1991, J. Phys. Chem., 95, 79,
  \dodoi{10.1021/j100154a019}

\bibitem[{{Calmonte} {et~al.}(2016){Calmonte}, {Altwegg}, {Balsiger},
  {Berthelier}, {Bieler}, {Cessateur}, {Dhooghe}, {van Dishoeck}, {Fiethe},
  {Fuselier}, {Gasc}, {Gombosi}, {H{\"a}ssig}, {Le Roy}, {Rubin}, {S{\'e}mon},
  {Tzou}, \& {Wampfler}}]{calmonte2016}
{Calmonte}, U., {Altwegg}, K., {Balsiger}, H., {et~al.} 2016, MNRAS, 462, S253,
  \dodoi{10.1093/mnras/stw2601}

\bibitem[{Caselli {et~al.}(1994)Caselli, Hasegawa, \& Herbst}]{caselli1994}
Caselli, P., Hasegawa, T., \& Herbst, E. 1994, ApJ, 421, 206,
  \dodoi{10.1086/173637}

\bibitem[{Cazaux {et~al.}(2022)Cazaux, Carrascosa, Caro, Caselli, Fuente,
  Navarro-Almaida, \& Rivi{\'e}re-Marichalar}]{cazaux2022}
Cazaux, S., Carrascosa, H., Caro, G.~M., {et~al.} 2022, A\&A, 657, A100,
  \dodoi{10.1051/0004-6361/202141861}

\bibitem[{Cioslowski {et~al.}(2001)Cioslowski, Szarecka, \&
  Moncrieff}]{cioslowski2001}
Cioslowski, J., Szarecka, A., \& Moncrieff, D. 2001, J. Phys. Chem. A, 105,
  501, \dodoi{10.1021/jp003339g}

\bibitem[{Das {et~al.}(2018)Das, Sil, Gorai, Chakrabarti, \& Loison}]{das2018}
Das, A., Sil, M., Gorai, P., Chakrabarti, S.~K., \& Loison, J.~C. 2018, ApJS,
  237, 9, \dodoi{10.3847/1538-4365/aac886}

\bibitem[{Dovesi {et~al.}(2018)Dovesi, Erba, Orlando, Zicovich-Wilson,
  Civalleri, Maschio, Rérat, Casassa, Baima, Salustro, \&
  Kirtman}]{dovesi2018}
Dovesi, R., Erba, A., Orlando, R., {et~al.} 2018, WIREs Comput. Mol. Sci., 8,
  e1360, \dodoi{10.1002/wcms.1360}

\bibitem[{Eckert \& Steudel(2003)}]{eckert2003}
Eckert, B., \& Steudel, R. 2003, Molecular Spectra of Sulfur Molecules and
  Solid Sulfur Allotropes, ed. R.~Steudel (Berlin, Heidelberg: Springer Berlin
  Heidelberg), 31--98, \dodoi{10.1007/b13181}

\bibitem[{Escribano {et~al.}(2014)Escribano, Timón, Gálvez, Maté, Moreno, \&
  Herrero}]{escribano2014}
Escribano, R., Timón, V., Gálvez, O., {et~al.} 2014, Phys. Chem. Chem. Phys.,
  16, 16694, \dodoi{10.1039/C4CP01573H}

\bibitem[{Ferrante {et~al.}(2008)Ferrante, Moore, Spiliotis, \&
  Hudson}]{ferrante2008}
Ferrante, R.~F., Moore, M.~H., Spiliotis, M.~M., \& Hudson, R.~L. 2008, ApJ,
  684, 1210, \dodoi{10.1086/590362}

\bibitem[{Ferrero {et~al.}(2020{\natexlab{a}})Ferrero, Mart{\'i}nez-Bachs,
  Enrique-Romero, \& Rimola}]{ferrero2020atoms}
Ferrero, S., Mart{\'i}nez-Bachs, B., Enrique-Romero, J., \& Rimola, A.
  2020{\natexlab{a}}, in Computational Science and Its Applications -- ICCSA
  2020, ed. O.~Gervasi, B.~Murgante, S.~Misra, C.~Garau, I.~Ble{\v{c}}i{\'{c}},
  D.~Taniar, B.~O. Apduhan, A.~M. A.~C. Rocha, E.~Tarantino, C.~M. Torre, \&
  Y.~Karaca (Cham: Springer International Publishing), 553--560,
  \dodoi{10.1007/978-3-030-58814-4_41}

\bibitem[{Ferrero {et~al.}(2020{\natexlab{b}})Ferrero, Zamirri, Ceccarelli,
  Witzel, Rimola, \& Ugliengo}]{ferrero2020}
Ferrero, S., Zamirri, L., Ceccarelli, C., {et~al.} 2020{\natexlab{b}}, ApJ,
  904, 11, \dodoi{10.3847/1538-4357/abb953}

\bibitem[{Fuente {et~al.}(2017)Fuente, Goicoechea, Pety, Gal,
  Martín-Doménech, Gratier, Guzmán, Roueff, Loison, Caro, Wakelam, Gerin,
  Riviere-Marichalar, \& Vidal}]{fuente2017}
Fuente, A., Goicoechea, J.~R., Pety, J., {et~al.} 2017, ApJL, 851, L49,
  \dodoi{10.3847/2041-8213/aaa01b}

\bibitem[{{Fuente} {et~al.}(2023){Fuente}, {Rivi{\`e}re-Marichalar},
  {Beitia-Antero}, {Caselli}, {Wakelam}, {Esplugues}, {Rodr{\'\i}guez-Baras},
  {Navarro-Almaida}, {Gerin}, {Kramer}, {Bachiller}, {Goicoechea},
  {Jim{\'e}nez-Serra}, {Loison}, {Ivlev}, {Mart{\'\i}n-Dom{\'e}nech},
  {Spezzano}, {Roncero}, {Mu{\~n}oz-Caro}, {Cazaux}, \&
  {Marcelino}}]{fuente2023}
{Fuente}, A., {Rivi{\`e}re-Marichalar}, P., {Beitia-Antero}, L., {et~al.} 2023,
  \aap, 670, A114, \dodoi{10.1051/0004-6361/202244843}

\bibitem[{Gibb {et~al.}(2000)Gibb, Whittet, Schutte, Boogert, Chiar,
  Ehrenfreund, Gerakines, Keane, Tielens, van Dishoeck, \& Kerkhof}]{gibb2000}
Gibb, E.~L., Whittet, D. C.~B., Schutte, W.~A., {et~al.} 2000, ApJ, 536, 347,
  \dodoi{10.1086/308940}

\bibitem[{{Gondhalekar}(1985)}]{gondhalekar1985}
{Gondhalekar}, P.~M. 1985, \mnras, 217, 585, \dodoi{10.1093/mnras/217.3.585}

\bibitem[{Grimme {et~al.}(2010)Grimme, Antony, Ehrlich, \& Krieg}]{grimme2010}
Grimme, S., Antony, J., Ehrlich, S., \& Krieg, H. 2010, J. Chem. Phys., 132,
  154104, \dodoi{10.1063/1.3382344}

\bibitem[{{Hasegawa} {et~al.}(1992){Hasegawa}, {Herbst}, \&
  {Leung}}]{hasegawa1992}
{Hasegawa}, T.~I., {Herbst}, E., \& {Leung}, C.~M. 1992, \apjs, 82, 167,
  \dodoi{10.1086/191713}

\bibitem[{Hudson {et~al.}(2015)Hudson, Gerakines, \& Loeffler}]{hudson2015}
Hudson, R.~L., Gerakines, P.~A., \& Loeffler, M.~J. 2015, Phys. Chem. Chem.
  Phys., 17, 12545, \dodoi{10.1039/C5CP00975H}

\bibitem[{Jenkins(2009)}]{jenkins2009}
Jenkins, E.~B. 2009, ApJ, 700, 1299, \dodoi{10.1088/0004-637X/700/2/1299}

\bibitem[{Jim{\'e}nez-Escobar \& Caro(2011)}]{jimenez2011}
Jim{\'e}nez-Escobar, A., \& Caro, G.~M. 2011, A\&A, 536, A91,
  \dodoi{10.1051/0004-6361/201014821}

\bibitem[{{Jim{\'e}nez-Escobar} {et~al.}(2014){Jim{\'e}nez-Escobar}, {Mu{\~n}oz
  Caro}, \& {Chen}}]{jimenez2014}
{Jim{\'e}nez-Escobar}, A., {Mu{\~n}oz Caro}, G.~M., \& {Chen}, Y.~J. 2014,
  \mnras, 443, 343, \dodoi{10.1093/mnras/stu1100}

\bibitem[{Laas \& Caselli(2019)}]{laas2019}
Laas, J.~C., \& Caselli, P. 2019, A\&A, 624, A108,
  \dodoi{10.1051/0004-6361/201834446}

\bibitem[{Lee {et~al.}(1988)Lee, Yang, \& Parr}]{lee1988}
Lee, C., Yang, W., \& Parr, R.~G. 1988, Phys. Rev. B, 37, 785,
  \dodoi{10.1103/PhysRevB.37.785}

\bibitem[{Lenain {et~al.}(1988)Lenain, Picquenard, Corset, Jensen, \&
  Steudel}]{lenain1988}
Lenain, P., Picquenard, E., Corset, J., Jensen, D., \& Steudel, R. 1988,
  Berichte der Bunsengesellschaft für physikalische Chemie, 92, 859,
  \dodoi{10.1002/bbpc.198800210}

\bibitem[{{Luna} {et~al.}(2017){Luna}, {Luna-Ferr{\'a}ndiz}, {Mill{\'a}n},
  {Domingo}, {Mu{\~n}oz Caro}, {Santonja}, \& {Satorre}}]{luna2017}
{Luna}, R., {Luna-Ferr{\'a}ndiz}, R., {Mill{\'a}n}, C., {et~al.} 2017, \apj,
  842, 51, \dodoi{10.3847/1538-4357/aa7562}

\bibitem[{Maschio {et~al.}(2013{\natexlab{a}})Maschio, Kirtman, Rérat,
  Orlando, \& Dovesi}]{maschio2013a}
Maschio, L., Kirtman, B., Rérat, M., Orlando, R., \& Dovesi, R.
  2013{\natexlab{a}}, J. Chem. Phys., 139, 164101, \dodoi{10.1063/1.4824442}

\bibitem[{Maschio {et~al.}(2013{\natexlab{b}})Maschio, Kirtman, Rérat,
  Orlando, \& Dovesi}]{maschio2013b}
---. 2013{\natexlab{b}}, J. Chem. Phys., 139, 164102, \dodoi{10.1063/1.4824443}

\bibitem[{McCarthy {et~al.}(2004{\natexlab{a}})McCarthy, Thorwirth, Gottlieb,
  \& Thaddeus}]{mccarthy2004_s3}
McCarthy, M.~C., Thorwirth, S., Gottlieb, C.~A., \& Thaddeus, P.
  2004{\natexlab{a}}, Journal of the American Chemical Society, 126, 4096,
  \dodoi{10.1021/ja049645f}

\bibitem[{McCarthy {et~al.}(2004{\natexlab{b}})McCarthy, Thorwirth, Gottlieb,
  \& Thaddeus}]{mccarthy2004_s4}
---. 2004{\natexlab{b}}, J. Chem. Phys., 121, 632, \dodoi{10.1063/1.1769372}

\bibitem[{McClure {et~al.}(2023)McClure, Rocha, Pontoppidan, Crouzet, Chu,
  Dartois, Lamberts, Noble, Pendleton, Perotti, {et~al.}}]{mcclure2023}
McClure, M.~K., Rocha, W. R.~M., Pontoppidan, K.~M., {et~al.} 2023, Nat.
  Astron., 7, 431–443, \dodoi{10.1038/s41550-022-01875-w}

\bibitem[{McElroy {et~al.}(2013)McElroy, {Walsh, C.}, {Markwick, A. J.},
  {Cordiner, M. A.}, {Smith, K.}, \& {Millar, T. J.}}]{umist}
McElroy, D., {Walsh, C.}, {Markwick, A. J.}, {et~al.} 2013, A\&A, 550, A36,
  \dodoi{10.1051/0004-6361/201220465}

\bibitem[{Meyer \& Stroyer-Hansen(1972)}]{meyer1972infrared}
Meyer, B., \& Stroyer-Hansen, T. 1972, J. Phys. Chem., 76, 3968,
  \dodoi{10.1021/j100670a013}

\bibitem[{{Minissale} {et~al.}(2022){Minissale}, {Aikawa}, {Bergin}, {Bertin},
  {Brown}, {Cazaux}, {Charnley}, {Coutens}, {Cuppen}, {Guzman}, {Linnartz},
  {McCoustra}, {Rimola}, {Schrauwen}, {Toubin}, {Ugliengo}, {Watanabe},
  {Wakelam}, \& {Dulieu}}]{minissale2022}
{Minissale}, M., {Aikawa}, Y., {Bergin}, E., {et~al.} 2022, ACS Earth Space
  Chem., 6, 597, \dodoi{10.1021/acsearthspacechem.1c00357}

\bibitem[{Pascale {et~al.}(2004)Pascale, Zicovich-Wilson, López~Gejo,
  Civalleri, Orlando, \& Dovesi}]{pascale2004}
Pascale, F., Zicovich-Wilson, C.~M., López~Gejo, F., {et~al.} 2004, J. Comput.
  Chem., 25, 888, \dodoi{10.1002/jcc.20019}

\bibitem[{Penteado {et~al.}(2017)Penteado, Walsh, \& Cuppen}]{penteado2017}
Penteado, E.~M., Walsh, C., \& Cuppen, H.~M. 2017, ApJ, 844, 71,
  \dodoi{10.3847/1538-4357/aa78f9}

\bibitem[{Perrero {et~al.}(2022)Perrero, Enrique-Romero, Ferrero, Ceccarelli,
  Podio, Codella, Rimola, \& Ugliengo}]{perrero2022}
Perrero, J., Enrique-Romero, J., Ferrero, S., {et~al.} 2022, ApJ, 938, 158,
  \dodoi{10.3847/1538-4357/ac9278}

\bibitem[{{Perrero} {et~al.}(2023){Perrero}, {Ugliengo}, {Ceccarelli}, \&
  {Rimola}}]{perrero2023}
{Perrero}, J., {Ugliengo}, P., {Ceccarelli}, C., \& {Rimola}, A. 2023, \mnras,
  525, 2654, \dodoi{10.1093/mnras/stad2459}

\bibitem[{Picquenard {et~al.}(1993)Picquenard, Boumedien, \&
  Corset}]{picquenard1993}
Picquenard, E., Boumedien, M., \& Corset, J. 1993, J. Mol. Struct., 293, 63,
  \dodoi{10.1016/0022-2860(93)80015-N}

\bibitem[{Raghavachari {et~al.}(1990)Raghavachari, Rohlfing, \&
  Binkley}]{raghavachari1990}
Raghavachari, K., Rohlfing, C.~M., \& Binkley, J.~S. 1990, J. Chem. Phys., 93,
  5862, \dodoi{10.1063/1.459583}

\bibitem[{Rauscher \& Ressler(2005)}]{rauscher2005jswt}
Rauscher, B.~J., \& Ressler, M.~E. 2005, Experimental Astronomy, 19, 149,
  \dodoi{10.1007/s10686-005-9015-0}

\bibitem[{Ruffle {et~al.}(1999)Ruffle, Hartquist, Caselli, \&
  Williams}]{ruffle1999}
Ruffle, D.~P., Hartquist, T.~W., Caselli, P., \& Williams, D.~A. 1999, MNRAS,
  306, 691, \dodoi{10.1046/j.1365-8711.1999.02562.x}

\bibitem[{Shingledecker {et~al.}(2020)Shingledecker, Lamberts, Laas, Vasyunin,
  Herbst, Kästner, \& Caselli}]{shingledecker2020}
Shingledecker, C.~N., Lamberts, T., Laas, J.~C., {et~al.} 2020, ApJ, 888, 52,
  \dodoi{10.3847/1538-4357/ab5360}

\bibitem[{Spitzer \& Jenkins(1975)}]{spitzer1975ultraviolet}
Spitzer, L., \& Jenkins, E.~B. 1975, ARA\&A, 13, 133,
  \dodoi{10.1146/annurev.aa.13.090175.001025}

\bibitem[{Steudel {et~al.}(2003)Steudel, Steudel, \& Wong}]{steudel2003}
Steudel, R., Steudel, Y., \& Wong, M.~W. 2003, Speciation and Thermodynamics of
  Sulfur Vapor, ed. R.~Steudel (Berlin, Heidelberg: Springer Berlin
  Heidelberg), 117--134, \dodoi{10.1007/b12405}

\bibitem[{Thorwirth {et~al.}(2005)Thorwirth, McCarthy, Gottlieb, Thaddeus,
  Gupta, \& Stanton}]{thorwirth2005}
Thorwirth, S., McCarthy, M.~C., Gottlieb, C.~A., {et~al.} 2005, J. Chem. Phys.,
  123, 054326, \dodoi{10.1063/1.1942495}

\bibitem[{{Tinacci} {et~al.}(2023){Tinacci}, {Germain}, {Pantaleone},
  {Ceccarelli}, {Balucani}, \& {Ugliengo}}]{tinacci2023}
{Tinacci}, L., {Germain}, A., {Pantaleone}, S., {et~al.} 2023, \apj, 951, 32,
  \dodoi{10.3847/1538-4357/accae8}

\bibitem[{{Tinacci} {et~al.}(2022){Tinacci}, {Germain}, {Pantaleone},
  {Ferrero}, {Ceccarelli}, \& {Ugliengo}}]{tinacci2022}
---. 2022, ACS Earth and Space Chemistry, 6, 1514,
  \dodoi{10.1021/acsearthspacechem.2c00040}

\bibitem[{Trofimov {et~al.}(2009)Trofimov, Sinegovskaya, \&
  Gusarova}]{trofimov2009}
Trofimov, B., Sinegovskaya, L., \& Gusarova, N. 2009, J. Sulfur Chem., 30, 518,
  \dodoi{10.1080/17415990902998579}

\bibitem[{Wakelam {et~al.}(2004)Wakelam, Castets, Ceccarelli, Lefloch, Caux, \&
  Pagani}]{wakelam2004}
Wakelam, V., Castets, A., Ceccarelli, C., {et~al.} 2004, A\&A, 413, 609,
  \dodoi{10.1051/0004-6361:20031572}

\bibitem[{Wakelam {et~al.}(2017)Wakelam, Loison, Mereau, \&
  Ruaud}]{wakelam2017}
Wakelam, V., Loison, J.-C., Mereau, R., \& Ruaud, M. 2017, Mol. Astrophys., 6,
  22 , \dodoi{10.1016/j.molap.2017.01.002}

\bibitem[{Wakelam {et~al.}(2015)Wakelam, Loison, Herbst, Pavone, Bergeat,
  B{\'{e}}roff, Chabot, Faure, Galli, Geppert, Gerlich, Gratier, Harada,
  Hickson, Honvault, Klippenstein, Picard, Nyman, Ruaud, Schlemmer, Sims,
  Talbi, Tennyson, \& Wester}]{kida}
Wakelam, V., Loison, J.-C., Herbst, E., {et~al.} 2015, ApJS, 217, 20,
  \dodoi{10.1088/0067-0049/217/2/20}

\bibitem[{Zamirri {et~al.}(2019)Zamirri, Macià~Escatllar, Mariñoso~Guiu,
  Ugliengo, \& Bromley}]{zamirri2020}
Zamirri, L., Macià~Escatllar, A., Mariñoso~Guiu, J., Ugliengo, P., \&
  Bromley, S.~T. 2019, ACS Earth Space Chem., 3, 2323,
  \dodoi{10.1021/acsearthspacechem.9b00157}

\bibitem[{Zicovich-Wilson {et~al.}(2004)Zicovich-Wilson, Pascale, Roetti,
  Saunders, Orlando, \& Dovesi}]{zicovich2004}
Zicovich-Wilson, C.~M., Pascale, F., Roetti, C., {et~al.} 2004, J. Comput.
  Chem., 25, 1873, \dodoi{10.1002/jcc.20120}

\end{thebibliography}
\bibliographystyle{aasjournal}

\appendix
\setcounter{table}{0}
\renewcommand{\thetable}{A\arabic{table}}

\section{S$_n$ species details}\label{appendix:species}

We computed the energy gaps between different spin states of the S$_n$ species characterized in this work. In the case of S and S$_2$, their ground state is a triplet spin state. The singlet and the triplet spin states are well separated in energy (see Table \ref{tab:energy_gaps_12}).

\begin{table}[htb]
    \centering
    \begin{tabular}{lcc}
         Species & Closed Shell Singlet & Triplet  \\
         \hline
         S & 431 & 0 \\
         S$_2$ & 274 & 0 \\
          \hline
    \end{tabular}
    \caption{Energy gaps (in kJ mol$^{-1}$) of atomic and diatomic sulphur species. The most stable structure is taken as the reference zero energy structure. }
    \label{tab:energy_gaps_12}
\end{table}     

In contrast, S$_3$, S$_4$, and S$_5$ are more elusive and present electronic states closed in energy (see Table \ref{tab:energy_gaps_345}). The ground state of S$_3$ and S$_4$ is an open shell singlet. For S$_3$ the closed shell chain structure is only 10 kJ mol$^{-1}$ less stable than the ground state, followed closely by the ring (in the close shell singlet state). In the case of S$_4$, the second most stable configuration is the triplet spin state chain. The cyclic form of S$_4$ is the less favoured conformer. S$_5$ most stable conformed has a ring structure and a singlet ground state. This species has also been modeled as a chain, however, it is almost 70  kJ mol$^{-1}$ less stable than the ring, when in the state of triplet or open shell singlet. The closed shell singlet configuration is remarkably unfavoured due to its planar geometry.

\begin{table*}[htb]
    \centering
    \begin{tabular}{lcccc}
         Species & Closed Shell Singlet & Open Shell Singlet & Triplet & Ring  \\
         \hline
         S$_3$ & 10 & 0 & 101 & 17\\
         S$_4$ & 43 & 0 & 21 & 60 \\
         S$_5$ &  162 & 68 & 67 & 0\\
         \hline
    \end{tabular}
    \caption{Energy gaps (in kJ mol$^{-1}$) of sulphur allotropes computed in this work. The most stable structure is taken as the reference zero energy structure. The ring conformer possesses a closed shell singlet spin state. For S$_5$, the ring conformer consists in the envelope conformation.}
    \label{tab:energy_gaps_345}
\end{table*}

Finally, it is well established that S$_6$, S$_7$, and S$_8$ are ring structures with a closed shell singlet ground state, thus we do not provide additional details regarding their electronic spin states. 

\section{Calculation of the Binding Energies} \label{appendix:be}

In the following, we report the detailed calculation of the binding energies.

We are interested in is the counterpoise-deformation-corrected interaction energy, where the basis set superposition error (BSSE) correction is included to make up for the error that arises when using a finite basis set of localized Gaussian functions to describe a chemical system:
\begin{equation}
   \begin{aligned}
    \Delta E^{CP} & = \Delta E^* + \delta E_S + \delta E_\mu + \delta E_L - BSSE \\
    & = \Delta E - BSSE
\end{aligned} 
\end{equation}

From this equation it appears that the non-counterpoise-corrected interaction energy $\Delta E$ is given by the sum of the deformation-free interaction energy ($\Delta E^*$), the deformation energy of the slab ($\delta E_S$) and the molecule ($\delta E_{\mu}$) and the lateral interaction ($E_L$) between adsorbate molecules in different replica of the cell. This quantity corresponds to the common definition of interaction energy, which is the difference between the energy of the complex and the energies of the isolated species and isolated ice model. 
\begin{equation}
   \begin{aligned}\label{eqn:deltaE}
    \Delta E & = \Delta E^* + \delta E_S + \delta E_{\mu} + \delta E_L \\
    & = E_{complex} - E_{ice} - E_{species}
\end{aligned} 
\end{equation}

The lateral interaction energy $\delta E_L$ is due to the interactions experienced by each adsorbate with its images on the replicated unit cells. This term becomes particularly relevant for high adsorbate coverage, as the distances between adsorbate molecules in neighbour cells will become shorter with the increasing coverage. Here, we are interested in minimizing the $\delta E_L$ term, to arrive to a $\Delta E$ representative of the sole interaction between the adsorbate and the ice surface. This requires the adoption of a large enough unit cell to minimize the $\delta E_L$ term.

By convention, the binding energy is simply the opposite of $\Delta$E$^{CP}$:
\begin{equation}
BE = -\Delta E^{CP}
\end{equation}

Therefore, the BE(0), that is the adsorption enthalpy at 0 K, can be obtained by simply subtracting the ZPE correction to the binding energy. In this work, we adopted $\Delta$ZPE = ZPE$_{complex}$ - ZPE$_{species}$, where ZPE$_{complex}$ was computed only for a fragment of the system constituted by the adsorbed species. In this way we are considering the same atoms in the two ZPE terms.

\begin{equation} \label{eqn:ZPE}
   BE(0)= BE - \Delta ZPE 
\end{equation}

\end{document}